\documentclass{article}

\usepackage{arxiv}
\usepackage[utf8]{inputenc}
\usepackage[T1]{fontenc}
\usepackage[colorlinks=true,allcolors=black]{hyperref}
\usepackage{url}
\usepackage{booktabs}
\usepackage{amsfonts}
\usepackage{nicefrac}
\usepackage{microtype}
\usepackage{graphicx}
\usepackage[numbers]{natbib}
\usepackage{doi}
\usepackage{xcolor}
\usepackage{amsmath}
\usepackage{array}
\usepackage{tabularx}
\usepackage{tikz}
\usepackage{pgfplots}
\usepgfplotslibrary{statistics}
\pgfplotsset{compat=1.18}
\usepackage{colortbl}
\usepackage{multirow}

\definecolor{posgreen}{RGB}{0,140,0}
\definecolor{negred}{RGB}{0,0,0}
\definecolor{posred}{RGB}{0,0,0}
\definecolor{fsmgray}{RGB}{250,250,200}

\begin{document}

% -----------------------------------------------

\title{Evaluating Interactivity:\\Toward Automated Assessment of AI-Generated Explorable Explanations}

\author{
Xiaozao Wang\\
New York University Shanghai\\
\texttt{xw2194@nyu.edu}\textsuperscript{*}
\And
Zhewei Wang\\
New York University Shanghai\\
\texttt{zw3636@nyu.edu}\thanks{Equal contribution.}
\And
Hongyi Wen\\
New York University Shanghai\\
\texttt{hw3242@nyu.edu}\thanks{Correspondence.}
}

\date{}

\maketitle

% -------------------- Abstract --------------------
\begin{abstract}
While large language models now enable rapid generation of interactive learning materials, evaluating the interaction quality of these \textit{explorable explanations} remains an open challenge. Existing benchmarks largely focus on code executability or visual fidelity, providing limited insight into dynamic interaction behaviors such as learner-controlled state transitions and context-sensitive system responses, which are factors that critically shape learners' conceptual understanding. We present \textbf{EE-Eval}, an automated evaluation framework that formalizes interactivity as a finite space of learner-controllable states and transitions, represented as a Finite State Machine (FSM). By extracting FSMs from AI-generated explorable explanations, EE-Eval externalizes implicit interaction logic into an explicit, machine-interpretable graph. Evaluation is performed by comparing each generated FSM to an ideal FSM that encodes pedagogical intent, using a combination of graph-based metrics and embedding-based comparison of states, actions, and feedback to measure their structural and semantic similarity. Across thousands of generated explorable explanations spanning 127 concepts and produced by 6 AI models, EE-Eval consistently differentiates interaction quality beyond surface-level criteria such as functional correctness or visual quality, and exhibits substantially stronger alignment with human judgments of interactivity and pedagogical effectiveness than existing baselines. By framing interactivity as testable behavioral models rather than an emergent byproduct of LLM generation, EE-Eval transforms evaluation into a reflective diagnostic tool, enabling pedagogically grounded and actionable human-AI collaboration in creating interactive educational content.


\end{abstract}

% =========================================================
\section{Introduction}

Explorable explanations have reshaped how complex computational concepts are taught and understood, shifting learning from passive consumption to active engagement through interaction. By enabling learners to manipulate parameters, observe system behaviors, and test hypotheses, explorable explanations foster intuition-building that static explanations often cannot, leading to their adoption in various domains. Recent advances in generative AI further enable the automatic generation of explorable explanations from high-level specifications or natural language prompts, including interactive visualizations and learning interfaces \cite{chen2025interactivesketchpadmultimodaltutoring,chen2022nl2interfaceinteractivevisualizationinterface,huang2025metaanalysisllmeffectsstudents}. Systems such as Chat2VIS \cite{maddigan2023chat2visgeneratingdatavisualisations}, ChartLlama \cite{han2023chartllamamultimodalllmchart}, and MatPlotAgent \cite{yang2024matplotagentmethodevaluationllmbased} illustrate this growing capability.

However, evaluating the quality of AI-generated explorable explanations remains an open challenge. Existing approaches primarily assess static outputs such as screenshots, textual summaries, or rendered charts using benchmarks like nvBench \cite{luo2021nvbenchlargescalesynthesizeddataset} and VisEval \cite{chen2024visevalbenchmarkdatavisualization}. While effective for measuring visual correctness and data fidelity, these methods fail to capture interaction dynamics and pedagogical effectiveness, which are central to learning outcomes.

Evaluating explorable explanations differs from assessing static educational content. Effectiveness depends not only on visual or textual fidelity, but on the interaction structure itself: what states learners can reach, what actions are available at each state, and how actions affect system behavior. Yet this interaction logic is implicit, embedded in low-level implementation details such as JavaScript event handlers and DOM manipulation \cite{Chulpongsatorn_2023,Diwan_2022}, making it difficult for humans and automated systems to reason about or compare interactive behaviors.

To address this gap, we propose \textbf{EE-Eval}, a framework that makes interaction logic explicit by representing explorable explanations as finite state machines (FSMs). This abstraction preserves interaction semantics while hiding implementation details. Each FSM encodes (1) meaningful interaction states, (2) user-triggered and conditional transitions, and (3) interface components involved in these transitions, yielding a shared and interpretable structure for analysis and benchmarking purposes \cite{gunturu2024augmentedphysicscreatinginteractive,guo2025agentsamastateawaremobileassistant,ku2025theoremexplainagentvideobasedmultimodalexplanations}.

Building on this representation, EE-Eval introduces structural and semantic similarity metrics to compare AI-generated explanations with ground truth references at the level of interaction behavior. Rather than evaluating whether explanations look similar, EE-Eval assesses whether they support comparable exploration and reasoning patterns that are essential for learning. We integrate a testing pipeline to measure coverage of states, transitions, and interface elements at scale. Experiments across algorithmic explorable explanations generated by multiple AI systems show that FSM-based evaluation yields more stable and informative signals than visual-based or execution-driven baselines, particularly in capturing interactivity and pedagogical structure, while producing interpretable diagnostics that complement human judgment.

Our contributions are summarized as follows:
\begin{enumerate}
\item We introduce an FSM-based representation that explicitly models the interaction logic of explorable explanations with a machine-interpretable graph.
\item We propose interaction-level structural and semantic metrics for evaluating AI-generated explorable content that aligns with human judgements.
\item We demonstrate how automated evaluation can function as a collaborative analytic tool for educators and researchers, supporting reflective assessment and iterative improvement of AI-generated learning materials.
\end{enumerate}

% =========================================================

\section{Related Work}

\subsection{Interactive Learning Systems}

Interactive learning systems emphasize learning through active exploration, hypothesis testing, and feedback. Prior work on interactive simulations and explorable explanations shows that direct manipulation and immediate system response can significantly improve conceptual understanding and engagement \cite{Banda:2023aa,Firat2018TowardsAS,kefalis_2025,Seskir_2022}. Empirical studies across STEM domains demonstrate that learners benefit more from interacting with dynamic models than from passively consuming static representations \cite{Banda:2023aa,kefalis_2025,kestin_2025}.

Systems such as TensorFlow Playground\footnote{\url{https://playground.tensorflow.org/}}, CNN Explainer \cite{Wang_2021}, Augmented Math \cite{Chulpongsatorn_2023}, and Augmented Physics \cite{gunturu2024augmentedphysicscreatinginteractive} exemplify this paradigm by allowing users to manipulate parameters and observe state changes in real time. These systems tightly couple user actions with system feedback, reinforcing learning-by-doing and exploratory reasoning. However, their interaction structures are manually designed and embedded implicitly in code, limiting scalability, adaptability, and systematic analysis of interaction behavior.

\subsection{AI-Generated Interactive Learning Materials}

Advances in large language models have shifted educational interfaces from manual authoring toward AI-generated learning materials. Meta-analyses report growth in systems that generate explanations, exercises, visualizations, and adaptive learning pathways \cite{bdcc9090237,Yan_2023}. Empirical studies show engagement and efficiency in AI-generated tutoring and content delivery \cite{Diwan_2022,kestin_2025}, while XR-based platforms integrate generative models into learning environments \cite{gianni_2025,zhang_2024}.

Within visualization and interface generation, prior work explores natural language–driven pipelines such as Chat2VIS \cite{maddigan2023chat2visgeneratingdatavisualisations}, ChartLlama \cite{han2023chartllamamultimodalllmchart}, NVAgent \cite{ouyang2025nvagentautomateddatavisualization}, and NL2INTERFACE \cite{chen2022nl2interfaceinteractivevisualizationinterface}. Agent-based and multimodal systems extend these ideas to pedagogical settings, including EduVisAgent \cite{ji2025eduvisbencheduvisagentbenchmarkmultiagent}, TheoremExplainAgent \cite{ku2025theoremexplainagentvideobasedmultimodalexplanations}, Interactive Sketchpad \cite{chen2025interactivesketchpadmultimodaltutoring}, Auto-Slides \cite{yang2025autoslidesinteractivemultiagentcreating}, and ReactGenie \cite{Yang_2024}, demonstrating that LLMs can generate executable interactive interfaces.

However, interaction is treated as an assumed outcome rather than a modeled object. Generated interfaces are presumed to function correctly, with limited mechanisms for verification or comparison. As interactivity shifts from a design-time artifact to a generation-time result, the lack of a machine-readable and verifiable representation becomes a limitation, particularly in educational contexts where correctness, controllability, and learner agency are critical \cite{ku2025theoremexplainagentvideobasedmultimodalexplanations,huang2025metaanalysisllmeffectsstudents,Yan_2023}.

\subsection{Evaluation for AI-Generated Learning Interfaces}

As AI systems generate interactive learning interfaces, evaluation has not kept pace with this shift. Most existing methods assess artifacts as outputs or isolated UI events, rather than modeling interaction as a state-driven process. Static-output benchmarks such as nvBench \cite{luo2021nvbenchlargescalesynthesizeddataset} and frameworks including VisEval \cite{chen2024visevalbenchmarkdatavisualization}, MatPlotAgent \cite{yang2024matplotagentmethodevaluationllmbased}, Visualization Generation with LLMs \cite{wang2025visualizationgenerationlargelanguage}, and Vi(E)va LLM \cite{podo2024vievallmconceptualstack} focus on executability, structural correctness, and visual quality, but evaluate only outputs and cannot capture interactive exploration. Sequential and video-based approaches, such as TheoremExplainAgent \cite{ku2025theoremexplainagentvideobasedmultimodalexplanations} and EduVisAgent \cite{ji2025eduvisbencheduvisagentbenchmarkmultiagent}, extend evaluation to temporal coherence, yet operate on pre-rendered sequences and exclude user-driven interaction and branching. Evaluation of interactive systems remains limited: UI testing tools simulate isolated events, and even when interactivity is considered, as in EduVisBench \cite{ji2025eduvisbencheduvisagentbenchmarkmultiagent}, it is treated as a set of discrete actions rather than a behavior model. Consequently, current methods cannot reason about interaction paths, state transitions, or the alignment between user actions and pedagogical intent in interactive learning systems.

% =========================================================
\section{FSM-based Evaluation of Interactivity}
Finite State Machines (FSMs) and state-based models have been used to represent interactive behavior in other domains. Mesbah et al. \cite{mesbah_2012} infer FSMs of AJAX-based web applications to support crawling and testing. In dialogue systems and intelligent tutoring, FSM-based planning structures interaction flow and long-term behavior \cite{WOO200625}, including in recent LLM-based frameworks such as FiSMiness \cite{zhao2025fisminessfinitestatemachine} and MAPLE \cite{guo2025agentsamastateawaremobileassistant}. These works show that explicit state modeling enables interpretability, executability, and automated verification.

Despite this success, state-based representations have rarely been applied to evaluating interactive visualizations or AI-generated learning interfaces. Consequently, current approaches lack a machine-readable and executable abstraction that captures user action–system response relationships and supports systematic testing and comparison. This gap motivates our proposal to model interactivity as a finite state machine for evaluating AI-generated explorable explanations.

Our work \textbf{EE-Eval} introduces an evaluation framework that conceptualizes interactivity as a structured space of learner-controllable states and transitions. Rather than treating interactivity as a collection of surface-level UI features, we model it as a state-transition system and evaluate its capacity, coherence, and pedagogical meaningfulness. FSMs serve as the core representational and analytical abstraction throughout our methodology.

\subsection{Representing Explorable Explanations with FSM}

An FSM externalizes interaction logic by explicitly encoding semantic states, user-initiated and conditional transitions, and observable system responses. By mapping how user actions lead to changes in system state, this abstraction reframes evaluation from unstructured code inspection to the comparison of state-transition structures that capture the essential aspects of interaction.

\begin{figure}[t]
\centering
\includegraphics[width=\textwidth]{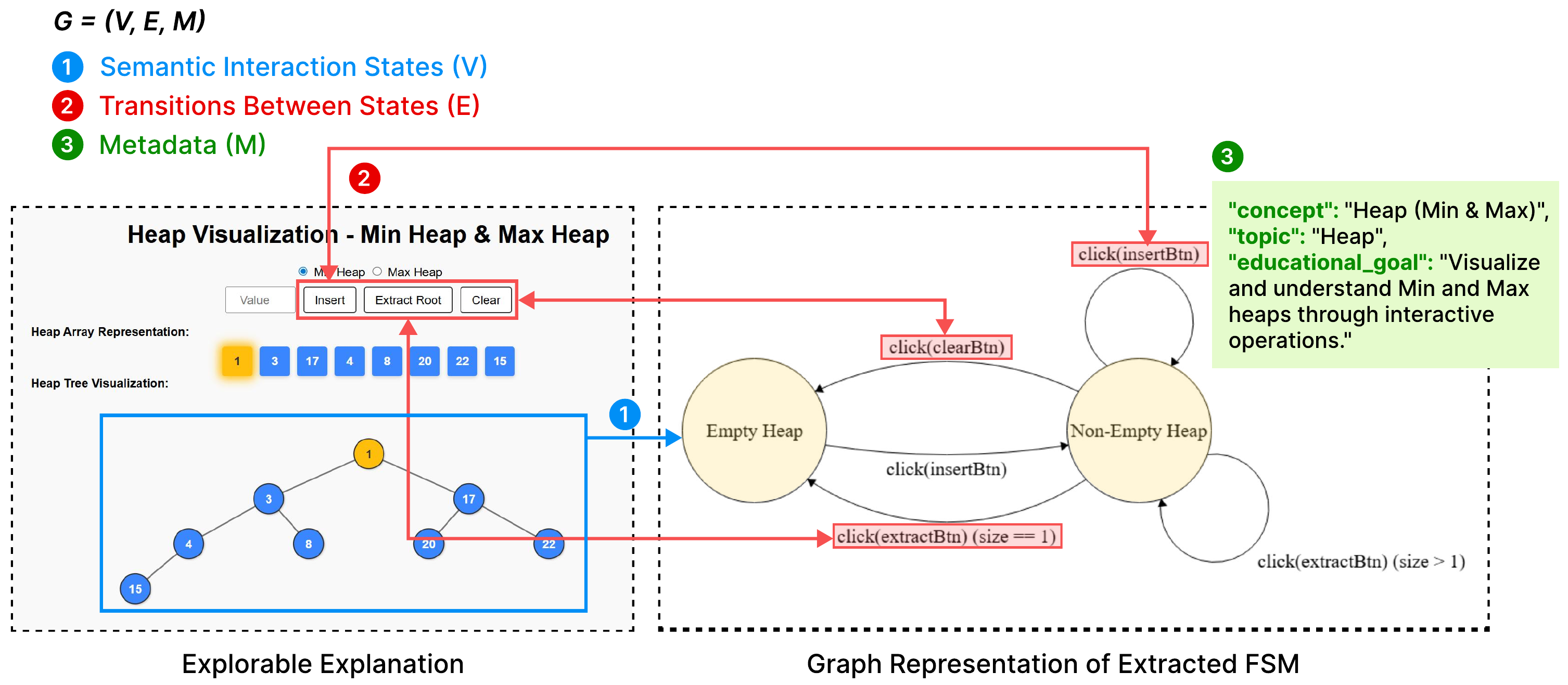}
\caption{Explorable Explanation and the Graph Representation of Extracted FSM.}
\label{fig:representation}
\end{figure}

Formally, each FSM is represented as a directed graph
\[
G = (V, E, M),
\]
where $V$ denotes semantic interaction states, $E$ denotes transitions between states, and $M$ captures task-level metadata, including pedagogical intent, target concepts, and variable definitions (Fig.~\ref{fig:representation}).

We operationalize FSMs using a JSON-based schema that captures the essential components of interactive behavior. Each JSON object contains metadata describing the educational goal and target concept, a set of \texttt{states} with entry and exit actions and observable UI cues, \texttt{events} representing user or system triggers, \texttt{transitions} linking source and target states via events, and \texttt{components} mapping abstract interaction elements to concrete DOM selectors.

This representation captures the essence of interactivity by explicitly modeling how learner actions propagate through the system, enabling systematic analysis of state coverage, transition coherence, and interaction correctness, while normalizing heterogeneous implementations for robust cross-model comparison.

\subsection{FSM Extraction from AI-Generated HTML}

To extract FSM representations from AI-generated interactive visualizations, we adopt a few-shot LLM-based strategy~\cite{brown2020languagemodelsfewshotlearners}. Due to the complexity and variability of HTML and JavaScript implementations, rule-based parsing is insufficient to recover high-level interaction semantics. We therefore provide the model with a small set of manually validated HTML–FSM pairs that demonstrate how interaction logic in DOM elements, event handlers, and scripts maps to our JSON-based FSM schema. Conditioned on these examples, the model extracts FSMs from unseen HTML inputs. The extracted FSMs are normalized and validated against schema constraints, including state uniqueness, transition consistency, and referential integrity, yielding structured and comparable representations for automatic evaluations. We qualitatively validated extracted FSMs on a subset of samples, confirming that major interaction states and transitions are consistently recovered.

\subsection{Evaluation Framework}

To quantitatively evaluate the quality of interaction structure in an extracted FSM, we measure its similarity to a pre-defined reference FSM.

Ideal FSMs are constructed through a semi-automated pipeline. Domain experts first define a set of pedagogically motivated design principles—such as explicit state progression, meaningful learner control, and coherent feedback loops—and create a small set of high-quality reference FSMs for representative concepts. Based on these examples, additional ideal FSMs are generated using a large language model to scale across topics, followed by targeted human verification to ensure structural validity and consistency. The ideal FSMs serve as consistent and interpretable reference structures that encode intended interaction patterns, enabling comparative evaluation across generated explanations.

\begin{figure}[t]
    \centering
    \includegraphics[width=1\linewidth]{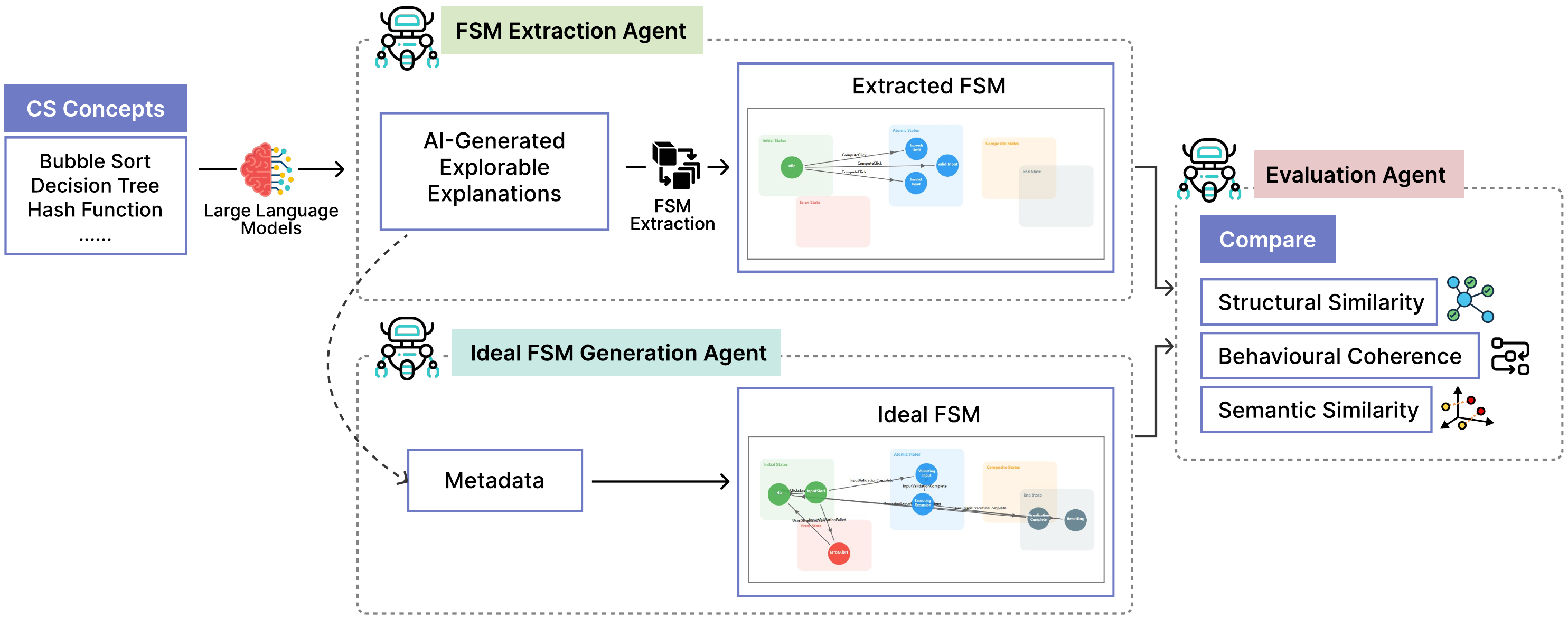}
    \caption{EE-Eval: FSM-based Multi-Agent Evaluation Framework.}
    \label{fig:framework}
% \vspace{-20pt}
\end{figure}

We propose a multi-dimensional evaluation framework capturing \emph{structural}, \emph{semantic}, and \emph{behavioral} similarity between the extracted and ideal FSMs. FSMs are normalized into a unified representation, and similarity is computed along these dimensions, which are integrated within a multi-agent evaluation pipeline (Fig.~\ref{fig:framework}). The final evaluation score is a weighted combination of these metrics.

\subsubsection{Structural Similarity}
Structural similarity captures alignment in graph topology, reflecting interaction
complexity and control-flow patterns. The most essential components we retain are node
and edge counts, which penalize over- or under-generation of states or transitions:

\[
S_{\text{node}} =
1 - \frac{\lvert |V_1| - |V_2| \rvert}{\max(|V_1|, |V_2|, 1)},
\quad
S_{\text{edge}} =
1 - \frac{\lvert |E_1| - |E_2| \rvert}{\max(|E_1|, |E_2|, 1)}.
\]

These metrics are complemented by higher-level structural assessments, such as degree
distributions and graph density. The structural similarity score aggregates these components into a single measure
$S_{\text{struct}}$, summarizing how closely the overall interaction structure of one FSM matches another.

\subsubsection{Semantic Similarity}

Structural alignment alone is insufficient to capture pedagogical or behavioral equivalence.

We therefore compute embedding-based semantic similarity on four axes: state labels,
actions, events, and FSM-level metadata. We use pre‑trained sentence embeddings from \texttt{sentence-transformers/all-MiniLM-L6-v2}, a compact embedding model based on the MiniLM architecture \cite{Wang2020MiniLM}, to compute semantic similarity between FSM elements.
 For example, for
state labels:

\[
\mathbf{s}_1 = \frac{1}{|L_1|} \sum_{\ell \in L_1} \text{Embed}(\ell),
\quad
\mathbf{s}_2 = \frac{1}{|L_2|} \sum_{\ell \in L_2} \text{Embed}(\ell),
\quad
S_{\text{state}} = \cos(\mathbf{s}_1, \mathbf{s}_2),
\]

and similarly for actions, events, and metadata. This approach ensures semantic
alignment even when lexical choices differ (e.g., ``idle'' vs.\ ``waiting''). The
overall semantic similarity $S_{\text{sem}}$ combines these four axes into a unified
metric.

\subsubsection{Behavioral Coherence}

To measure whether interactions unfold consistently, we compute a FSM isomorphism score $S_{\text{iso}}$. Derived from a canonical signature of node degrees and semantic categories, $S_{\text{iso}}$ equals $1$ for exact matches and $0$ otherwise. This metric directly captures behavioral coherence: FSMs with identical structural and semantic signatures inherently produce the same state-transition sequences, ensuring that user actions elicit consistent system responses. Future work may explore graded structural similarity measures.

\subsubsection{Combined Similarity Score}

Finally, we integrate all components into a weighted aggregate similarity score:

\[
S_{\text{total}} =
\alpha \cdot S_{\text{struct}} +
\beta \cdot S_{\text{sem}} +
\gamma \cdot S_{\text{iso}},
\]

where \(\alpha, \beta, \gamma \in [0,1]\) and \(\alpha + \beta + \gamma = 1\).  
This weighting balances structural fidelity, semantic alignment, and exact behavioral correspondence. The weights \(\alpha, \beta, \gamma\) are determined through a constrained grid search over a set of candidate combinations. Specifically, we evaluate different weight configurations and select \((0.4, 0.4, 0.2)\), which yields the strongest alignment with human ratings on a validation subset.

% =========================================================
\subsection{Automated Evaluation Pipeline and Interface}

We implement the proposed evaluation framework within an automated multi-agent pipeline
that integrates generation, extraction, and analysis. Our code is available at GitHub Repository
\href{https://github.com/WangZhewei1027/EE-Eval}{\textbf{EE-Eval}}. The pipeline first generates
interactive explanations using multiple AI models, extracts FSMs from the resulting
HTML/JavaScript artifacts, and computes FSM-based evaluation scores across all dimensions.

\begin{figure}[t]
\centering
\begin{minipage}{0.49\textwidth}
  \centering
  \includegraphics[width=\linewidth]{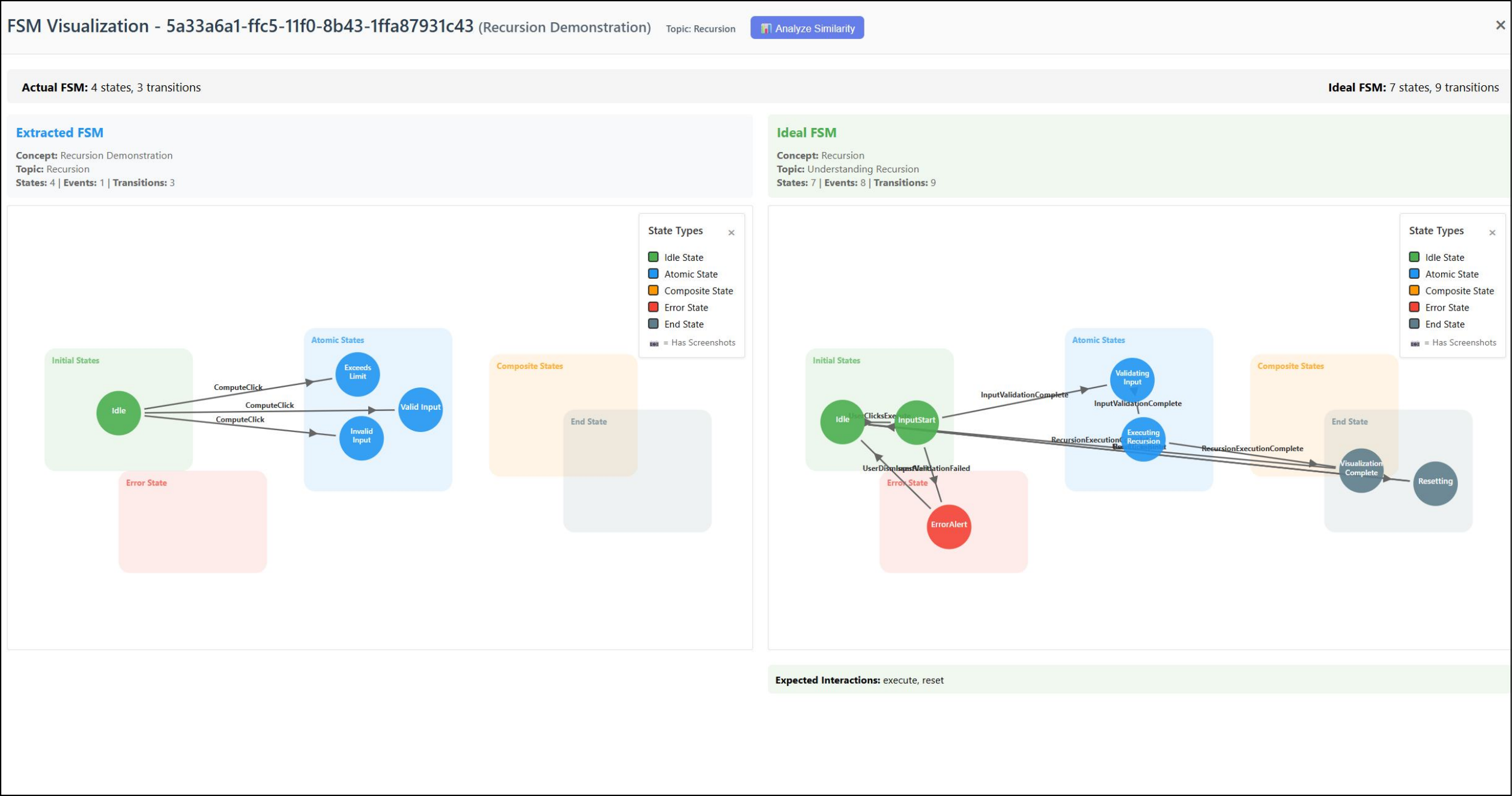}
  \\ (a) FSM Evaluation interface
  \label{fig:ui-left}
\end{minipage}\hfill
\begin{minipage}{0.49\textwidth}
  \centering
  \includegraphics[width=\linewidth]{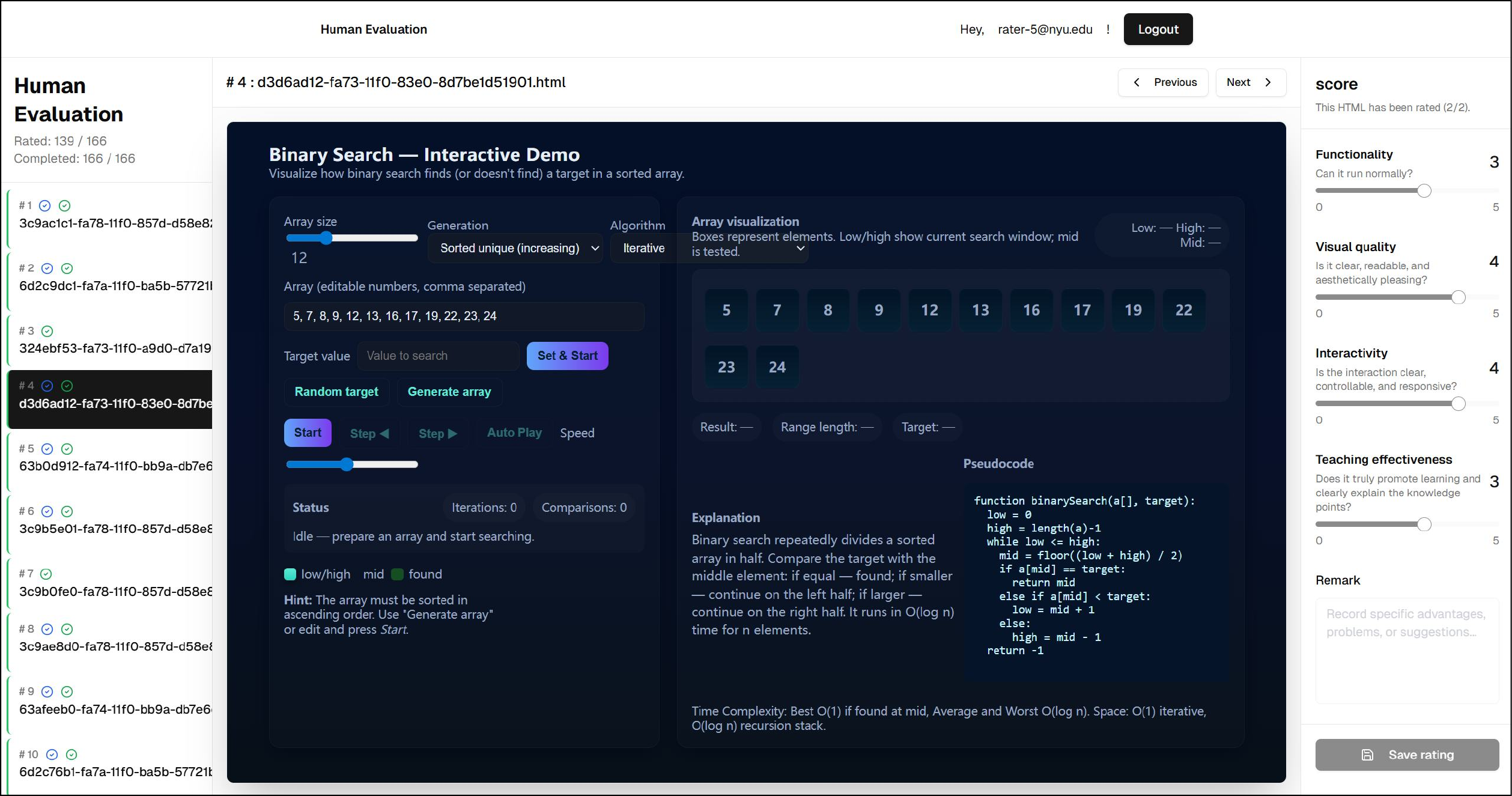}
  \\ (b) Human evaluation interface
  \label{fig:ui-right}
\end{minipage}
\caption{EE-Eval's user-facing interfaces}
\label{fig:dashboard}
\end{figure}

To support transparency and interpretability, we provide a user-facing visualization
interface that allows inspection of generated UIs, corresponding FSM structures, and
detailed evaluation results (Fig.~\ref{fig:dashboard}). This interface enables both automated large-scale analysis
and qualitative inspection of interaction logic, facilitating error diagnosis and
comparative evaluation.

% =========================================================
\section{Results and Discussion}
\subsection{Experimental Setup}
To evaluate the effectiveness of our FSM-based framework, we conducted an empirical study on 2497 AI-generated explorable explanations covering 127 computer science concepts. We tested six models of varying capabilities, including three closed-source models (GPT-3.5 Turbo\footnote{\href{https://openai.com/blog/chatgpt}{OpenAI GPT-3.5 Turbo}},
GPT-4o-mini\footnote{\href{https://openai.com/index/gpt-4o-mini-advancing-cost-efficient-intelligence/}{OpenAI GPT-4o Mini}},
GPT-5 Mini\footnote{\href{https://arxiv.org/abs/2601.03267}{OpenAI GPT-5 Mini}})
and three open-source models
(DeepSeek-V3\footnote{\href{https://arxiv.org/abs/2412.19437}{DeepSeek-V3}},
Qwen1.5-0.5B\footnote{\href{https://arxiv.org/abs/2309.16609}{Qwen1.5-0.5B-Chat}},
Llama-3.2-1B\footnote{\href{https://arxiv.org/abs/2407.21783}{Llama-3.2-1B}}).
Our analysis addresses two central questions:

\textbf{RQ1:} Does FSM-based evaluation capture interaction quality that aligns with human judgment, beyond surface-level visual or functional cues emphasized by existing baselines?

\textbf{RQ2:} Given such alignment, can FSM-based evaluation reliably discriminate interactive quality across models and content types?

\subsubsection{Human Evaluation as Ground Truth}
To assess alignment with human judgment, we conducted a controlled study on a randomly sampled subset of 166 explorable explanations. All six raters have formal training in computer science and prior experience with interactive learning systems, enabling informed and consistent judgment. Each explanation was independently rated by two raters on four dimensions: \textit{Functional correctness, Visual quality, Interactivity, and Pedagogical effectiveness}. Raters were briefed on the criteria and shown calibration examples before evaluation. All samples were presented in randomized order under a blind setup with no information about the generating model. Raters demonstrated reasonable consistency in their relative judgments across evaluation dimensions. Final ratings for each explorable explanation were averaged across raters to produce a single score per dimension.

\subsection{Baselines}
We compare \textbf{EE-Eval} against two representative baselines. The \textbf{Unit Test} baseline uses LLM-generated Playwright\footnote{\url{https://playwright.dev/}} tests to automatically exercise UI components and verify executability on the DOM level. The \textbf{VLM} (Vision-Language Model) baseline, implemented with GPT-4o-mini, evaluates visual quality by scoring screenshots from initial and terminal states, reflecting perception-based assessment without access to interaction behavior. Together, these baselines cover execution- and appearance-driven evaluation practices, providing contrast to FSM-based evaluation.

\subsection{Alignment with Human Judgment (RQ1)}

A key requirement for any evaluation metric of interactive systems is that it reflects how humans perceive interaction quality. Without such alignment, automated metrics may overemphasize superficial features while failing to capture meaningful interaction behavior. We first examine whether FSM-based evaluation captures interaction quality in a manner consistent with human judgment, particularly in contrast to baselines that emphasize surface-level visual appearance or isolated functional execution. This question is critical because interactive learning quality is defined not by how an interface looks or whether individual actions execute, but by how user actions lead to coherent, pedagogically meaningful state transitions.

To disentangle interaction quality from surface-level cues, we evaluate alignment under two settings. The \textit{balanced} dataset reflects the natural distribution of generated explorable explanations, while the \textit{biased} dataset is constructed to amplify specific surface features (e.g., strong visual presentation but weak interaction structure). This separation allows us to test whether evaluation methods remain sensitive to interaction behavior when visual cues become misleading.

\begin{table}[t]
\centering
\label{tab:pearson-correlation-combined}
\renewcommand{\arraystretch}{1.3}

\resizebox{\textwidth}{!}{%
\begin{tabular}{>{\bfseries}l >{\bfseries}l c c c c c}
\toprule
Data & Framework & Functional & Visual & Interactivity & Pedagogical & Overall \\
\midrule

\rowcolor{fsmgray}
\multirow{3}{*}{\textbf{(a) Balanced}}
& FSM-based 
& \textcolor{posred}{\textbf{0.676}$^{***}$}
& \textcolor{posred}{\textbf{0.690}$^{***}$}
& \textcolor{posgreen}{\textbf{0.728}$^{***}$}
& 0.580$^{***}$
& \textcolor{posgreen}{\textbf{0.696}$^{***}$} \\

& VLM (Baseline 1)
& 0.537$^{***}$
& \textcolor{posred}{\textbf{0.628}$^{***}$}
& 0.530$^{***}$
& 0.557$^{***}$
& 0.584$^{***}$ \\

& Unit test (Baseline 2)
& \textcolor{negred}{-0.580}$^{***}$
& \textcolor{negred}{-0.630}$^{***}$
& \textcolor{negred}{-0.600}$^{***}$
& \textcolor{negred}{-0.492}$^{***}$
& \textcolor{negred}{-0.598}$^{***}$ \\

\midrule

\rowcolor{fsmgray}
\multirow{3}{*}{\textbf{(b) Biased}}
& FSM-based
& 0.065
& \textcolor{negred}{-0.148}
& \textcolor{posgreen}{\textbf{0.210}}
& 0.125
& 0.079 \\

& VLM (Baseline 1)
& 0.178
& \textcolor{posred}{\textbf{0.674}$^{***}$}
& -0.019
& 0.001
& \textcolor{posred}{\textbf{0.242}$^{*}$} \\

& Unit test (Baseline 2)
& \textcolor{negred}{-0.189}
& \textcolor{negred}{-0.298$^{**}$}
& -0.090
& 0.029
& \textcolor{negred}{-0.162} \\

\bottomrule
\end{tabular}
}
\vspace{0.5em}
\caption{Pearson Correlation ($r$) between Automated Frameworks and Human Evaluation. 
\textbf{(a)} Balanced Data. \textbf{(b)} Biased Data. 
\textbf{\textit{Note:}} $^{*}p<0.05$, $^{**}p<0.01$, $^{***}p<0.001$.}

\end{table}

We first analyze alignment on the balanced dataset, which reflects the natural distribution of AI-generated explorable explanations (Table~\ref{tab:pearson-correlation-combined}(a)). Across all four dimensions, FSM-based evaluation demonstrates the strongest overall alignment with human judgment. Notably, FSM achieves the highest correlation on \textbf{Interactivity} ($r=0.728, p<0.001$), substantially outperforming VLM-based evaluation ($r=0.530$) and unit testing ($r=-0.600$). This result directly supports our central hypothesis: explicitly modeling interaction as a state machine enables the evaluator to capture interaction behaviors that humans perceive but surface-oriented metrics overlook.

FSM also exhibits strong alignment on \textbf{Functional} ($r=0.676$) and \textbf{Visual} ($r=0.690$), resulting in the highest overall correlation ($r=0.696$). In contrast, VLM-based evaluation aligns most strongly with Visual quality, reflecting its reliance on rendered appearance rather than executable interaction logic. Unit testing shows consistently negative correlations, suggesting that brittle test scripts disproportionately penalize complex but high-quality interactive designs. Together, these results indicate that, under realistic conditions, baseline methods are systematically influenced by surface cues, whereas FSM-based evaluation more faithfully reflects human assessments of interaction quality.

\subsubsection{Analysis on Biased Data}

We further evaluate alignment on a biased dataset, intentionally constructed to amplify specific surface features, such as strong visual presentation or fluent interaction with weak conceptual relevance (Table~\ref{tab:pearson-correlation-combined}(b)). Under these conditions, correlations decrease across all methods, reflecting the difficulty of disentangling interaction quality from dominant surface cues. Despite this shift, FSM-based evaluation retains a relative advantage on \textbf{Interactivity} ($r=0.210$), outperforming VLM ($r=-0.019$) and unit testing ($r=-0.090$). In contrast, VLM shows the strongest alignment on Visual quality ($r=0.674, p<0.001$), consistent with its appearance-driven assessment. These results highlight the complementary scope of the methods and reinforce the robustness of FSM-based evaluation for assessing interaction behavior under biased conditions.

\subsection{Discriminative Power of FSM-Based Evaluation (RQ2)}

\begin{figure}[t]
\centering
\includegraphics[width=1\textwidth]{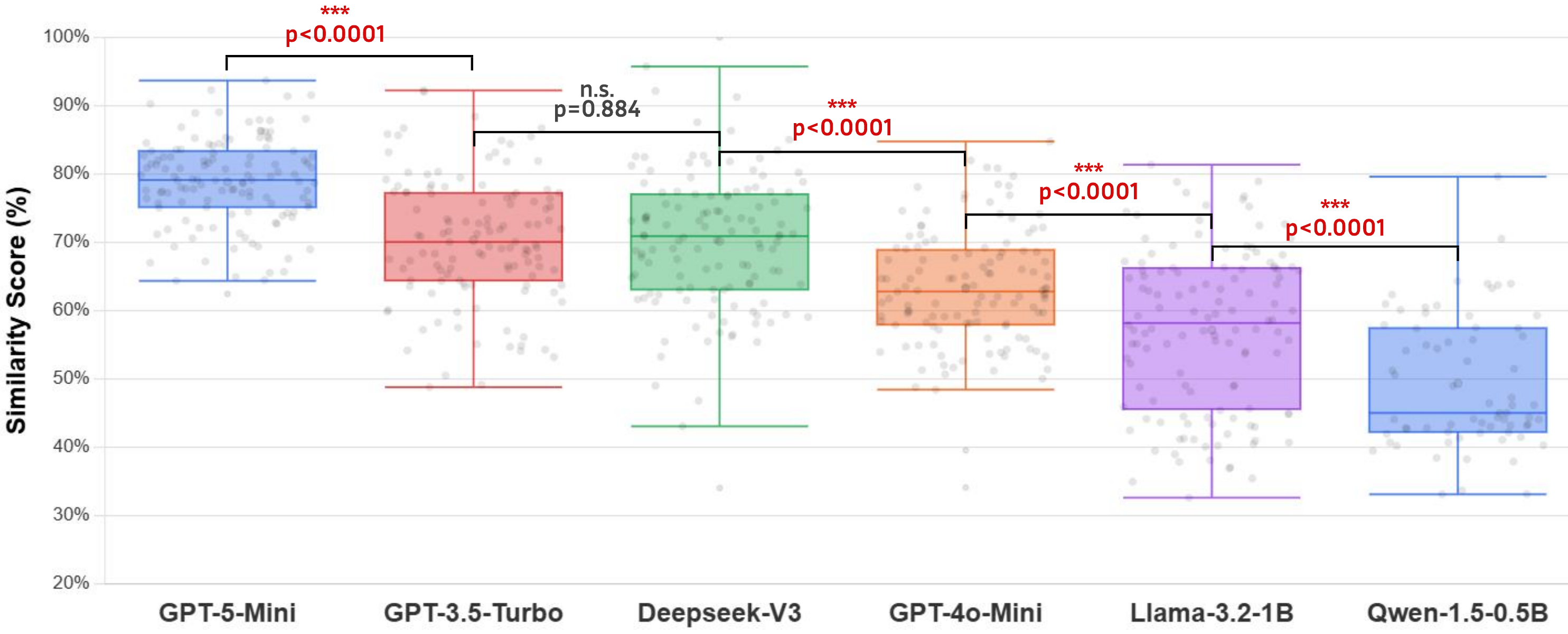}
\caption{FSM Similarity Score Distribution by Model}
\label{fig:distribution}
\end{figure}

Beyond alignment with human judgment, an effective evaluation metric should enable meaningful comparison across models and content types. Without sufficient discriminative power, evaluation results cannot support benchmarking or guide system improvement. Building on alignment with human judgment, we assess whether FSM-based evaluation discriminates model and content differences (Fig.~\ref{fig:distribution}). We report results from two-sample Student’s t-tests, where $\Delta\mu$ denotes the difference in mean scores between model pairs and $p$ indicates statistical significance. FSM metrics clearly separate models: GPT-5 Mini achieves the highest overall score (78.85\%) and behavioral coherence (33.13\%), significantly outperforming GPT-3.5-Turbo ($\Delta\mu = 8.54$, $p<0.0001$). GPT-3.5-Turbo and DeepSeek-V3 show no significant difference ($\Delta\mu = 0.18$, $p=0.884$), while subsequent tiers exhibit highly significant separations ($p<0.0001$). FSM evaluation further reveals content-dependent interaction demands: structured topics yield more stable transitions, whereas procedural topics show greater variance. Together, these results demonstrate that FSM-based evaluation robustly distinguishes model capabilities and captures task-specific interaction complexity beyond aggregate performance metrics.

Interestingly, our FSM-based evaluation diverges from coding-oriented leaderboards such as APXML\footnote{\url{https://apxml.com/leaderboards/coding-llms}}, where GPT-4o-Mini (Z-score 0.30) ranks above GPT-3.5-Turbo (Z-score -2.98) in aggregate reasoning and code generation ability. Under FSM metrics, however, GPT-3.5-Turbo consistently outperforms GPT-4o-Mini despite its lower nominal capacity. We attribute this reversal to FSM-based evaluation capturing latent interaction properties that are largely invisible to static or execution-based benchmarks. While GPT-4o-Mini often produces functionally correct interfaces, these outputs frequently under-specify interaction states, resulting in fragmented or ambiguous transitions. In contrast, GPT-3.5-Turbo more consistently externalizes interaction structure in a stable, analyzable form, yielding higher FSM coverage and behavioral coherence. This finding suggests that larger model capacity does not necessarily translate to stronger interaction quality, and that FSM-based evaluation exposes a complementary dimension of interactive competence beyond existing leaderboard metrics.

\subsubsection{Summary}

Our established FSM-based evaluation is both a valid and useful framework for assessing AI-generated interactive learning materials. By demonstrating strong alignment with human judgments on interactivity, FSM evaluation confirms that it captures interaction behavior beyond surface-level visual or functional cues. Building on this validity, its discriminative power across models and content types shows that such interaction-aware metrics enable meaningful comparison and diagnosis. These results position FSM-based evaluation not merely as a necessary abstraction for understanding, comparing, and improving AI-generated explorable explanations.

% =========================================================

\section{Discussion and Future Directions}

This study underscores a central challenge for AI-generated explorable explanations: existing evaluation approaches fail to account for interaction as a core learning mechanism. LLM-based static judgments, VLM-based screenshot scoring, and DOM-level unit tests largely overlook how learners act, explore, and receive feedback over time. As a result, these methods weakly align with human judgment and struggle to distinguish model capabilities, due to implicit interaction logic, unstable structures, and missing semantic representations. EE-Eval addresses this gap by making interaction behavior explicit through FSMs. Yielding interpretable and discriminative metrics that correlate with human ratings across more than 2,500 explanations spanning 127 computer science topics, EE-Eval shows stronger correlation with human ratings and provides interpretable signals about interaction complexity and coverage. Importantly, we interpret these signals as reflecting pedagogical coherence at the interaction level, rather than direct measures of learning outcomes.

More broadly, EE-Eval reframes interactivity as a testable artifact. By representing states, transitions, and component bindings explicitly, the framework supports fine-grained analysis of how learners and systems co-act, enabling clearer error diagnosis, cross-model comparison, and interpretable identification of structural and semantic weaknesses. In this sense, evaluation becomes not only analytic but also pedagogically informative, supporting the design of AI systems as reliable learning partners rather than opaque tools.

At the same time, our approach has several limitations. First, FSMs are an abstraction that may not fully capture highly continuous or emergent interaction patterns, potentially simplifying rich user experiences into discrete transitions. Second, the quality of evaluation depends on the accuracy of FSM extraction from generated artifacts, which may introduce noise or bias despite the use of few-shot prompting. Third, the construction of “ideal FSMs” involves domain assumptions about effective interaction design, which may not generalize across all topics or pedagogical contexts. Finally, EE-Eval focuses on interaction structure and does not assess other critical dimensions such as content correctness, visual clarity, or actual learning outcomes.

These limitations point to several directions for future work. One important direction is to connect interaction-level evaluation with downstream learning processes, for example through controlled user studies or proxy behavioral signals, to better understand how specific interaction patterns support understanding. Another direction is to improve FSM representations and extraction methods, potentially incorporating probabilistic transitions or hybrid symbolic–neural approaches to better capture complex interactions. In addition, FSM-based signals could be integrated into generation pipelines as structured feedback or reward signals, enabling models to iteratively improve interaction design rather than optimizing for static outputs alone.

Taken together, these considerations suggest that FSM-based evaluation should be viewed not as a replacement for existing methods, but as a complementary perspective that addresses their blind spots. While vision-based approaches are effective at assessing visual layout and perceptual clarity, FSM-based evaluation captures the underlying interaction logic that governs learner exploration. Integrating behavioral, structural, and perceptual signals may therefore offer a more complete and robust framework for evaluating AI-generated educational systems. In this sense, EE-Eval contributes a scalable and interpretable layer for assessing interaction quality, positioning interactivity as a first-class dimension in the evaluation of AI-supported learning environments.

% =========================================================
\section{Conclusion}
This work presented EE-Eval, an FSM-based framework for evaluating AI-generated explorable explanations by making interaction logic explicit through states, transitions, and component bindings. Moving beyond static evaluation, EE-Eval captures dynamic behaviors that shape learner experience and pedagogical effectiveness. Experiments across multiple generative models show that FSM-based metrics reliably distinguish model quality, align with human judgment, and reveal content-dependent interaction complexity. By treating evaluation as an analytic and pedagogical tool, EE-Eval provides interpretable diagnostics and actionable feedback, laying a foundation for iterative generation pipelines and more transparent, reliable, and human-centered AI-assisted educational systems.

\section*{Acknowledgments}
This project was supported in part through STCSM 23YF1430300 and NYU Shanghai Center for Data Science.

\bibliographystyle{plain}
\bibliography{references}

\end{document}